\documentclass[a4paper, 11pt]{article}

\usepackage{a4wide}
\usepackage{amsmath}
\usepackage{commath}
\usepackage{latexsym}
\usepackage{graphicx}
\usepackage{color}
\usepackage[pdfstartview=FitH]{hyperref}
\usepackage{verbatim}
\usepackage{authblk}
\usepackage{titling}
\usepackage[subtle]{savetrees}

\title{A stochastic differential equation approach to the analysis of the UK 2017 and 2019 general election polls}

\author{Mark Levene mark@dcs.bbk.ac.uk \authorcr}
\author{Trevor Fenner trevor@dcs.bbk.ac.uk}
\affil{Department of Computer Science and Information Systems, \authorcr Birkbeck, University of London, London WC1E 7HX, U.K.}

\date{}

\begin{document}

\maketitle

\begin{abstract}

Human dynamics and sociophysics build on statistical models that can shed light on and add to our understanding of social phenomena.
We propose a generative model based on a stochastic differential equation that enables us to model the opinion polls leading up to the UK 2017 and 2019 general elections, and to make predictions relating to the actual result of the elections. After a brief analysis of the time series of the poll results, we provide empirical evidence that the gamma distribution, which is often used in financial modelling, fits the marginal distribution of this time series. We demonstrate that the proposed poll-based forecasting model may improve upon predictions based solely on polls. The method uses the Euler-Maruyama method to simulate the time series, measuring the prediction error with the mean absolute error and the root mean square error, and as such could be used as part of a toolkit for forecasting elections.


\end{abstract}


\section{Introduction}

We propose to model election polls as a time series \cite{CHAT19}, motivated by \cite{WLEZ02}, who considered modelling the sequence of polls as an autoregressive process. Poll-based election forecasting methods \cite{FISH18}, building on vote-intention polls, play an important role in the endeavour to predict the outcome of an election; see Section~\ref{sec:prev} for a brief review on the methods of forecasting elections. We demonstrate that the model we propose, based on {\em stochastic differential equations} (SDEs) \cite{MACK11,EVAN13}, has the potential to give better predictions of the actual election result than simply using the results of the polls themselves.

\smallskip

In particular, we present a companion paper to \cite{FENN18a} using the same methodology, which is based on SDEs applied to opinion polls leading up to an election rather than to a referendum. We deploy a novel stochastic process based on the {\em Cox-Ingersoll-Ross} (CIR) process \cite{COX85,CHOU06}, used to model the term structure of interest rates \cite{BERK17}. CIR processes are `mean-reverting' diffusion processes \cite{HIRS14}, and have marginal distributions which are gamma distributed. Moreover, processes that are sums of such diffusions have autocorrelation functions (also known as serial correlation functions) that are sums of the exponentially decaying autocorrelation functions of the constituent diffusion processes \cite{BIBB05,FORM08}, allowing the approximation of heavy tailed distributions \cite{FELD98}.

\smallskip

We refer the reader to \cite{FENN18a} for the background in human dynamics and sociophysics \cite{SEN14} (also known as {\em social physics}), noting that statistical physics \cite{CAST09} has played a central role in its formulation; humans are viewed as ``social atoms'', each exhibiting simple individual behaviour having limited complexity, but nevertheless collectively they yield complex social patterns \cite{LEVE19a}. In the context of human dynamics, the SDE model we propose is a {\em generative model} in the form of a stochastic process the evolution of which gives rise to distributions such as power law and Weibull distributions \cite{FENN15}.
Generative models also arise from {\em agent-based models} \cite{CONT14} and have played an important role in the sociophysics literature in the context of opinion dynamics \cite{CAST09,SIRB17}. In particular, the voter model and its extensions \cite{CAST09,SIRB17}, whereby at each time step an agent decides whether to hold on to or change its opinion depending on the opinions of its neighbours, have applications in explaining and understanding voting behaviour during elections.

\smallskip

Opinion polls, which provide the data source for our SDE model, relay important information to the public in the lead-up to an election and provide an important ingredient of forecasting methods; see \cite{TRAU16} for a high-level overview of election polls. In a given election cycle, polls can be naturally viewed as a time series, and thus be expected to follow a stochastic process, such as an autoregressive model of order 1 (or more succinctly an AR(1) model) \cite{CHAT19}. In \cite{WLEZ02} the authors had some reservations about using such a time series model, due to sampling error and lack of sufficient time series data, and thus proposed to analyse the data in terms of a time series cross-sectional model \cite{BECK11}, treating data as cross sections for each time unit in the election cycle. Furthermore, in \cite{WLEZ17} it was mentioned that, given a sufficient number of poll results, these could be readily treated as a statistical time series. The availability of a sufficient number of polls, in our case leading up to the 2017 and 2019 UK general elections, and a more general stochastic model, such as the one we propose, allow for the resurrection of poll-based forecasting using time series.

\smallskip

In \cite{FENN18a} we took a fresh look at the time series approach, going beyond the model suggested in \cite{WLEZ02}, and made use of the availability of a large number of polls conducted at regular intervals. In particular, we proposed a novel model based on SDEs, which are widely used in physics and mathematical finance to model diffusion processes, that can be viewed as continuous approximations to discrete processes modelling how the polls vary over time. Therein we provided empirical evidence that the beta distribution, which is a natural choice when modelling proportions, fits the marginal distribution of the time series and we provided evidence of the predictive power of the model (cf. \cite{KONO17,MORI19}). One disadvantage of this model is that its autocorrelation function is decreases exponentially \cite{BIBB05}, while in reality the tails of the autocorrelation function may be heavier. We address this problem in Section~\ref{sec:model} by extending the model of \cite{FENN18a} to allow processes that are sums of diffusions \cite{BIBB05,FORM08}, in which case the autocorrelation function is a sum of exponentials.

\smallskip

In order to evaluate the predictive power of the model, we make use of the {\em Euler-Maruyama} (EM) method \cite{SAUE13}, which is a computational method for approximating numerical solutions to SDEs. In particular, the EM method allows us to simulate the time series in order to predict the result of the election from the SDEs.
We utilise the well-known {\em mean absolute error} (MAE) and {\em root mean square error} (RMSE) metrics \cite{CHAI14} to assess the accuracy of the EM method in predicting the actual election result, and we compare these to the predictions obtained by simply taking the results of the opinion polls; see \cite{JENN20} for a discussion on the use MAE and RMSE  for assessing the forecasting performance of polls.

\smallskip

The rest of the paper is organised as follows.
In Section~\ref{sec:prev}, we review related research on election forecasting.
In Section~\ref{sec:polls}, we provide a brief analysis of the UK election poll results for 2017 and 2019.
In Section~\ref{sec:model}, we propose a generative model for analysing the polling data based on a sum of `mean-reverting' stochastic differential equations. In Section~\ref{sec:uk}, we apply the model to the polls leading up to the UK 2017 and 2019 general elections, utilising the EM method to evaluate the predictive power of the model. Finally, in Section~\ref{sec:conc}, we give our concluding remarks.

\section{A brief review of forecasting elections and previous research}
\label{sec:prev}

Forecasting election results focuses the mind on what is important in influencing election outcomes \cite{FISH18}. Its goal is clear, to predict which party will win the elections. Here will only consider two-party systems and, in particular, we examine the contest between the Conservative and Labour parties in the UK; however, the model we present is also relevant to other two-party systems such as in the USA, where the contest is between the Republicans and Democrats.  Although in election forecasting, the task at hand is to predict the winner, it is also about understanding how elections work and how effective the proposed models really are.

\smallskip

We now briefly outline the prominent election forecasting methods.
Structural models are based on fitting a regression model \cite{GELM20} to historical election data and using the results for prediction, assuming a causal relationship exists between the past and present. The independent variables, or predictors, are referred to as {\em fundamentals}, and most often include economic indicators (i.e. how did the economy perform) and leadership evaluations (i.e. how did the leaders perform).

\smallskip

As opposed to structural models, poll-based forecasting is based on voter intention. Two main challenges of poll-based forecasting are: how to aggregate polls and what model to use for the actual forecasting \cite{PASE15,JACK16}. Early approaches using time series to model polls for the purpose of building predictive models, were proposed by \cite{ERIK99} and \cite{GREE99}. Both of these models were especially concerned with reducing the sampling error of polls, and thereby with methods for smoothing the time series data.

\smallskip

Our focus is on the forecasting model itself rather than in aggregation, and to this end modelling the polls as a time series like in those early approaches mentioned above. However, we view the time series of polls as a diffusion process, which is a continuous approximation to discrete processes modelling the changes to polls over time. In particular, we propose to use `mean-reverting' diffusion processes arising from a particular class of SDEs \cite{BIBB05}, which describe CIR processes \cite{COX85}. Using the Euler-Maruyama method, mentioned in the introduction, the continuous SDEs of the CIR process are approximated by discrete processes analogous to the AR(1) process \cite{CHAT19}, as suggested in \cite{WLEZ02}. The model we propose, is however `mean reverting' and thus possesses a stationary solution. Moreover, the marginal distribution of the solution is a gamma distribution \cite{COBB81,BIBB05}, an aspect which is further discussed in Section~\ref{sec:model}.

\smallskip

It is important to note that polls are not a panacea for forecasting election outcomes, and they may fail to provide accurate predictions, as in the 2015 UK elections where the polling samples were unrepresentative of the target population's voting intentions \cite{STUR18}. We also mention that the model presented in \cite{FORD16}, which includes the aggregation of polls from various sources, taking into account historical polling data used to calibrate the prediction, and also provides, through simulation, UK constituency-level forecasts. Furthermore, it was demonstrated in \cite{WLEZ02,WLEZ17} that, as one would expect, polls are generally more accurate the closer they are in the election cycle to the actual election. Synthetic models \cite{LEWI16b}, and, more generally hybrid models \cite{PASE15} combine poll-based and structural models to obtain the advantages of both. One such example is the model of \cite{LINZ13}, which proposes a synthetic dynamic Bayesian model that provides both national-level and state-level forecasts.

\smallskip

Prediction markets provide another data source for forecasters, with the argument that in this case, since people are betting on the result, they will take all available information into account. However, it is not clear whether prediction markets perform better than polls. In \cite{READ19} the authors concluded that opinion polls are favourable in terms of their bias (the mean error of all forecasts), while prediction markets are better in terms of their precision (the reciprocal of the variance of all forecasts).

\smallskip

Citizen forecasting is the process of aggregating forecasts made by individuals, which can be viewed as a form of `wisdom of crowds'. While polls are based on voters intentions, citizen forecasting is based on voter expectations. In \cite{MURR19}, empirical evidence is provided that election forecasts based on voter expectation outperform those based on voter intention. In general, it would advisable to augment poll-based prediction with voter expectation surveys, should they become readily available. In principle, the techniques used for polls, as the one suggested herein, could be easily adapted to expectation surveys.
Furthermore, combining any number of the forecasting methods discussed, and weighting them according to their perceived accuracy, may lead to more accurate forecasts \cite{GRAE14}.

\smallskip

Another way to distinguish between forecasts, proposed by \cite{LEWI16a} is to contrast the {\em long view} with the {\em short view}. Taking the {\em long view}, performance is examined over several election contests and forecasts are made well before election day, while in the {\em short view} performance is measured iteratively and depends increasingly on polls as election day gets closer. Here we take the short view based solely on polls, however we note that synthetic models, which combine both views, can mitigate against inaccurate polling.

\smallskip

Another method, which has become popular due to the availability of social media data is {\em nowcasting} \cite{CERO17}, a method whose aim is to predict the present or the very near present, rather than the future. So, suppose that social media data are available, such as textual content from Twitter. Then, using this data, sentiment analysis \cite{LIU15} of the text can be computed, and, if it indicates a positive intention to vote for a particular party, this information can, in principle, be used in lieu of polling information. Moreover, since past Twitter data are available, they may also be combined into a time series and employed for forecasting, using, for example, the model we propose.

\section{Preliminary analysis of the time series of poll results}
\label{sec:polls}

\begin{figure}[ht]
\begin{center}
\includegraphics[scale=0.45]{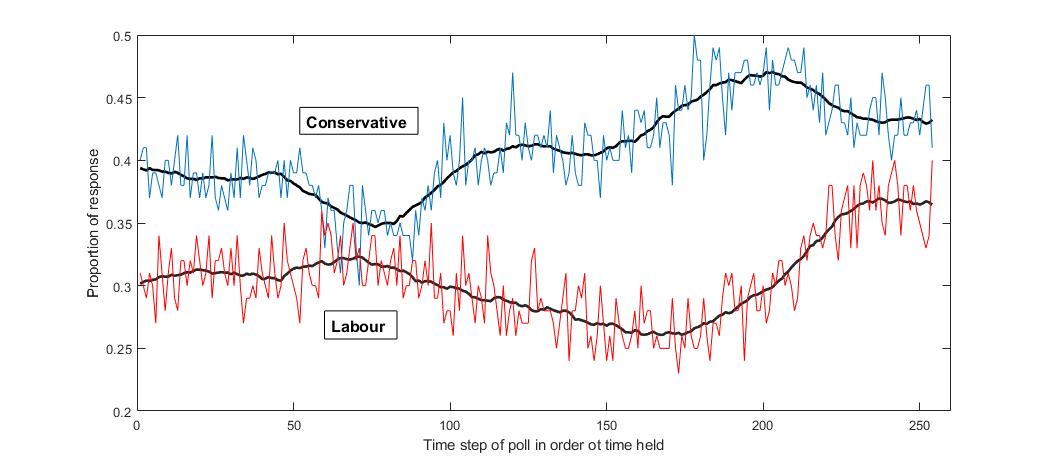}
\caption{\label{figure:ma17} Raw time series and moving averages of the 2017 polls with a centred sliding window of 25 time steps for the Conservative and Labour parties.}
\end{center}
\end{figure}

\begin{figure}[ht]
\begin{center}
\includegraphics[scale=0.4]{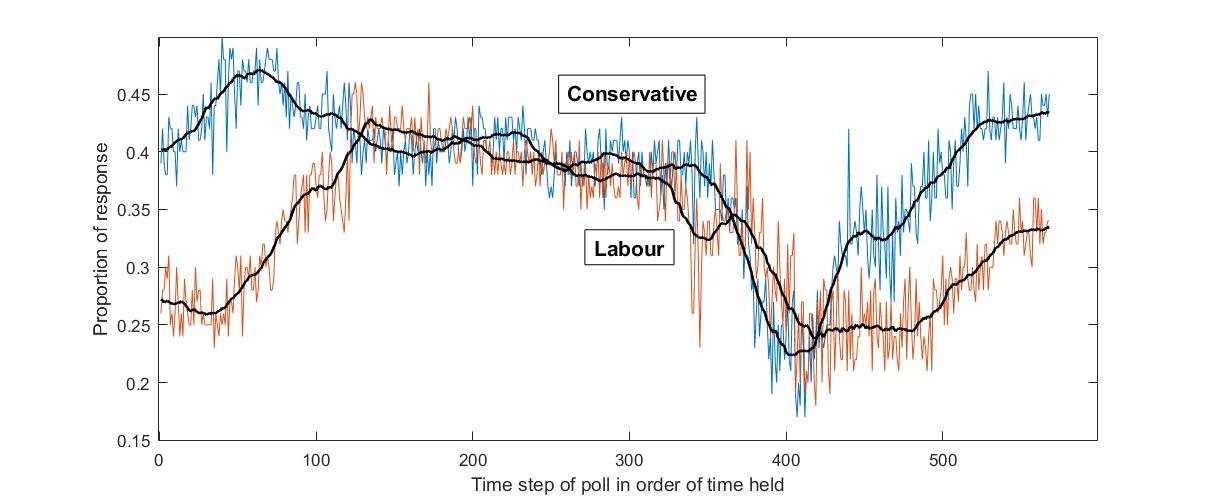}
\caption{\label{figure:ma19} Raw time series and moving averages of the 2019 polls with a centred sliding window of 25 time steps for the Conservative and Labour parties.}
\end{center}
\end{figure}

The analysis for the 2017 election was carried out using the results of 254 opinion polls, which were collected prior to the election that took place on 8th June 2017. The data set was obtained online from \cite{FT17}, the first poll being taken on 9th May 2015 and the last on the day before the election. Detailed results of the election can be obtained online from \cite{BBC17}. Similarly, the analysis for the 2019 election was carried out using the results of 568 opinion polls, which were collected prior to the election that took place on 12th December 2019. The data set was obtained online from \cite{FT19}, the first poll being taken on 4th January 2017 and the last on the day before the election. Detailed results of the election can be obtained online from \cite{BBC19}.
For each party and for each election, the data set used was a time series of the proportion of respondents who said they would vote for that party. These are shown graphically in Figures \ref{figure:ma17} and \ref{figure:ma19}; the minimum and maximum values of the intended vote share, for the Conservative and Labour parties, over all the 2017 and 2019 polls are shown in Tables~\ref{table:minmax-2017} and \ref{table:minmax-2019}, respectively.

\begin{table}[ht]
\begin{center}
\begin{tabular}{|l|c|c|}\hline
2017 elections & Min  & Max  \\ \hline \hline
Conservative   & 30\% & 50\% \\ \hline
Labour         & 23\% & 40\% \\ \hline
\end{tabular}
\end{center}
\caption{\label{table:minmax-2017} Minimum and maximum percentages of the intended vote share from the 2017 polls.}
\end{table}
\begin{table}[ht]
\begin{center}
\begin{tabular}{|l|c|c|}\hline
2019 elections & Min  & Max  \\ \hline \hline
Conservative   & 17\% & 50\%  \\ \hline
Labour         & 18\% & 46\%  \\ \hline
\end{tabular}
\end{center}
\caption{\label{table:minmax-2019}  Minimum and maximum percentages of the intended vote share from the 2019 polls.}
\end{table}
\smallskip

When analysing the data, in order to detect any clear trends, it is interesting to inspect the moving averages \cite{CHAT19} of the polls, which are shown in Figures \ref{figure:ma17} and \ref{figure:ma19}. For the 2017 election, it is clear that, although there was a dip in the support for Labour as the election was approaching, as it got closer to the election date the gap between Conservative and Labour narrowed, until the last day before the election when the Conservative lead in the polls was only 1\%. In the election itself, where the actual result was that the Conservatives received 42.4\% of the vote and Labour 40.0\%, the Conservative lead was slightly higher at 2.4\%. In 2019, the election date of 12th December was decided in parliament on the 29th October, and the Conservative lead in the polls from that date until the election was on average 10.8\%, with a standard deviation of 3.37\%. The Conservative lead on the last day before the elections was 11\% and in the election itself, in which the actual result was that the Conservatives received 43.6\% of the vote and Labour 32.2\%, the lead was even higher at 11.4\%. This does not tell the whole story of this election as the UK ``first-pass-the-post'' electoral system  resulted in the Conservatives ending up with a majority of 80 seats in parliament.

\smallskip

In our model and analysis given below we treat the two parties and two elections independently, with the realisation that in practice the time series for the Conservatives and Labour parties are not actually independent and that there may be dependencies between consecutive elections; we view our model and analysis as a first approximation to the poll-based forecasting problem. We note that although the two main parties in the UK receive most of the votes, there is at least a third party, the Liberal Democrats, which we have not considered in this analysis, but could be considered in future research. In this context, it is worth noting that there have been times when one of the two main parties receives more votes, possibly at the expense of the other (see Figure~\ref{figure:ma17}), and there are other times when both of the parties have received more votes, possibly at the expense of a third party (see Figure~\ref{figure:ma19}).


\section{A generative model for time series with application to polls}
\label{sec:model}

Stochastic differential equations can provide effective generative models for time series.
In particular, when the SDEs are `mean-reverting' \cite{HIRS14}, as will be the case here, they often possess stationary solutions that fit a number of well-studied distributions \cite{COBB81,BIBB05}. In our application, analysing the poll results, the gamma distribution \cite{JOHN94c,DAGP19} appears to be a natural choice, since it is flexible and allows the construction of a sum of diffusion processes having an autocorrelation function that is a sum of exponentials \cite{BIBB05,FORM08}.

\smallskip

A typical {\em stochastic differential equation} (SDE) takes the form
\begin{equation}\label{eq:sde}
{\mathrm d} X_t = \mu(X_t) {\mathrm d} t + \sigma(X_t) {\mathrm d} W_t,
\end{equation}
where $X_t$ is a random variable with $t \ge 0$ a real number denoting time; $\mu$ and $\sigma$ are known as the {\em drift} and {\em diffusion} functions, respectively, and $W_t$ is a Wiener process (also known as Brownian motion).
Moreover, when
\begin{equation}\label{eq:drift}
\mu(x) = \theta \left(m - x \right),
\end{equation}
where $\theta$, the {\em rate parameter}, is a positive constant and $m$ is a constant representing the mean of the underlying stochastic process, the SDE has a stationary solution \cite{COBB81,BIBB05}. In addition, its {\em autocorrelation function} is exponentially decreasing \cite{BIBB05} and takes the form
\begin{equation}\label{eq:autocorr}
\exp(- \theta t).
\end{equation}

Such a stochastic process is known as a `mean-reverting' process.
It was shown in \cite{COBB81,BIBB05} that, if
\begin{equation}\label{eq:diffusion}
\sigma^2(x) = \frac{2 \theta}{\lambda} x
\end{equation}
and $\lambda$ is defined by
\begin{equation}\label{eq:mean}
m = \frac{\alpha}{\lambda},
\end{equation}
the marginal distribution of the stationary solution of the SDE is a gamma distribution \cite{JOHN94c} with probability density function
\begin{equation}\label{eq:gamma}
\frac{\lambda^\alpha}{\Gamma(\alpha)} x^{\alpha-1}  \exp(- \lambda x),
\end{equation}
where $\Gamma$ is the gamma function \cite[6.1]{ABRA72}; $\alpha > 0$ is the {\em shape} of the distribution and $\lambda > 0$ is a {\em scale} parameter.
We note that several other forms for $m$ and $\sigma^2(X_t)$ also lead to well-known distributions \cite{COBB81,BIBB05}.

\medskip

Under the above conditions, substituting (\ref{eq:diffusion}) and (\ref{eq:mean}) into (\ref{eq:sde}) gives the SDE
\begin{equation}\label{eq:cir}
{\mathrm d} X_t =  \theta \left(\frac{\alpha}{\lambda} - X_t \right){\mathrm d} t + \sqrt{\frac{2 \theta}{\lambda} X_t} \ {\mathrm d} W_t,
\end{equation}
which describes a process called the {\em Cox-Ingersoll-Ross process} (CIR process) \cite{COX85}. From the results in \cite{FELL51} (cf. \cite{COX85}),
we can conclude that the solution to (\ref{eq:cir}) is positive when $\alpha \ge 1$.

\smallskip

The autocorrelation function of the CIR process is exponential as in (\ref{eq:autocorr}), and therefore, in order to model a process having an autocorrelation function with a heavier tail (such as a power law), we introduce a diffusion process that is a sum of CIR processes \cite{BIBB05,FORM08}, for which the autocorrelation function is a sum of exponentials. This relies on the result that a power law can be approximated by a finite sum of exponentials, since it is a completely monotone function \cite{FELD98} (cf. \cite{FENN16a}).

\smallskip

We can obtain a process $X_t$ as the sum of $n$ processes by letting
\begin{equation}\label{eq:sumi}
X_t = X_t^{(1)} + X_t^{(2)} + \cdots + X_t^{(n)},
\end{equation}
where the Wiener processes $W_t^{(i)}$ are independent, for $1 \le i \le n$, and $X_t^{(i)}$ is defined by the SDE
\begin{equation}\label{eq:cir_sumi}
{\mathrm d} X_t^{(i)} =  \theta_i \left(\frac{\phi_i \alpha}{\lambda} - X_t^{(i)} \right){\mathrm d} t + \sqrt{\frac{2 \theta_i}{\lambda} X_t^{(i)}} \ {\mathrm d} W_t^{(i)}.
\end{equation}
\smallskip

Then the mean, diffusion squared and autocorrelation function are given, respectively, by
\begin{equation}\label{eq:meani}
m_i = \frac{\phi_i \alpha}{\lambda},
\quad \sigma^2\left(X_t^{(i)}\right) = \frac{2 \theta_i}{\lambda} X_t^{(i)} \ \ {\rm and}
\quad \exp( - \theta_i t),
\end{equation}
where
\begin{equation}\label{eq:coeffi}
\phi_1 + \phi_2 + \cdots + \phi_n = 1.
\end{equation}
\smallskip

It follows that the marginal distribution of each $X_t^{(i)}$ is a gamma distribution with shape $\phi_i \alpha$ and scale parameter $\lambda$.
The marginal distribution of the sum $X_t$ is a gamma distribution with shape $\alpha$ and the same scale parameter $\lambda$.
Moreover, $X_t$ has autocorrelation function
\begin{equation}\label{eq:autocorr-sum}
\phi_1 \exp(- \theta_1 t) + \phi_2 \exp(- \theta_2 t) + \cdots + \phi_n \exp(- \theta_n t).
\end{equation}

\section{Analysis of the poll results for the general elections}
\label{sec:uk}

The approach we have taken to validate the model is similar to that taken in \cite{TAUF07}, building on the stationary diffusion-type models developed in \cite{BIBB05} for constructing diffusion processes with a given marginal distribution and autocorrelation function.

\smallskip

We can simulate the sum of the diffusion processes defined by (\ref{eq:sumi}) and (\ref{eq:cir_sumi}) using the {\em Euler-Maruyama} (EM)  method \cite{SAUE13} (cf. \cite{DERE12}), which is a general computational method for obtaining approximate numerical solutions to SDEs. We also make use of the {\em Jensen-Shannon divergence} ($JSD$) \cite{ENDR03} as a goodness-of-fit measure \cite{LEVE18}. All computations were carried out using the Matlab software package.

\smallskip

In Tables \ref{table:gamma17} and \ref{table:gamma19} we show the parameters of the gamma distributions fitted to the data sets using the maximum likelihood method, and the $JSD$ between the empirical marginal distribution of the time series of the poll results and the fitted gamma distribution; we note that its mean $\mu$ is given by $\mu = \alpha/\lambda$, and its standard deviation $\sigma$ by $\sigma^2 = \alpha/\lambda^2$. The low $JSD$ values indicate good fits for both political parties.
We note that the $JSD$ for the Conservative party in the 2019 elections is much higher than that for 2017, which could indicate that another distribution may better fit the data.
In fact, we found that the Gumbel distribution \cite{JOHN95b,KOTZ00} (also known as a type I extreme value distribution) is a better fit for the Conservatives in 2019, with $JSD$ of $0.0595$, although a worse fit for Labour in 2019  with a $JSD$ of $0.0801$.
We also note that, on inspection of Tables~\ref{table:minmax-2017} and \ref{table:minmax-2019}, it would seem that employing a truncated gamma distribution \cite{ZANI13b} would lead to a more accurate, albeit more complex, model. We leave these lines of investigation for future work as, for the purpose of prediction, the gamma distribution seems to be sufficient.

\begin{table}[ht]
\begin{center}
\begin{tabular}{|l|c|c|c|c|c|}\hline
2017 elections & $\alpha$ & $\lambda$  & $\mu$  & $\sigma$ & $JSD$  \\ \hline \hline
Conservative   & 105.8670  & 258.5847 & 0.4094 & 0.0398   & 0.0370 \\ \hline
Labour         & 72.0295   & 236.1929 & 0.3050 & 0.0359   & 0.0478 \\ \hline
\end{tabular}
\end{center}
\caption{\label{table:gamma17} Maximum likelihood fitting of the gamma distribution to the 2017 election polls.}
\end{table}
\begin{table}[ht]
\begin{center}
\begin{tabular}{|l|c|c|c|c|c|}\hline
2019 elections & $\alpha$ & $\lambda$  & $\mu$  & $\sigma$ & $JSD$  \\ \hline \hline
Conservative   & 32.0520  & 83.0583    & 0.3859 & 0.0682   & 0.1117 \\ \hline
Labour         &  25.2832 & 75.8911    & 0.3332   & 0.0663 & 0.0470 \\ \hline
\end{tabular}
\end{center}
\caption{\label{table:gamma19} Maximum likelihood fitting of the gamma distribution to the 2019 election polls.}
\end{table}
\smallskip

We fit the autocorrelation function in (\ref{eq:autocorr-sum}), for $n=2$, using least squares nonlinear regression in order to obtain estimates $\phi_i$ and $\theta_i$, for $i =1,2$. The results for the 2017 and 2019 poll results are shown in Tables~\ref{table:autocorr17} and \ref{table:autocorr19}, respectively, where we first smoothed the autocorrelations using a moving average with a window of length two. For comparison purposes we also show in Tables~\ref{table:autocorr17-exp1} and \ref{table:autocorr19-exp1}, fits for a single exponential, i.e. when $n=1$, for 2017 and 2019, respectively, which are worse than the fits for a mixture of two exponentials, i.e. when $n=2$.

\begin{table}[ht]
\begin{center}
\begin{tabular}{|l|c|c|c|c|c|}\hline
2017 elections & $\phi_1$ & $\theta_1$ & $\phi_2$ & $\theta_2$ & $JSD$ \\ \hline \hline
Conservative   & 0.8092   & 0.0146     & 0.1908   & 0.9896     & 0.0074 \\ \hline
Labour         & 0.7509   & 0.0237     & 0.2491   & 1.1890     & 0.0179\\ \hline
\end{tabular}
\end{center}
\caption{\label{table:autocorr17} Parameters of the exponential sum autocorrelation for the 2017 election polls.}
\end{table}
\begin{table}[ht]
\begin{center}
\begin{tabular}{|l|c|c|c|c|c|}\hline
2019 elections & $\phi_1$ & $\theta_1$ & $\phi_2$ & $\theta_2$ & $JSD$ \\ \hline \hline
Conservative   & 0.9717   & 0.0089     & 0.0283   & 2.3243     & 0.0035 \\ \hline
Labour         & 0.9439   & 0.0066     & 0.0561   & 1.1521     & 0.0025 \\ \hline
\end{tabular}
\end{center}
\caption{\label{table:autocorr19} Parameters of the exponential sum autocorrelation for the 2019 election polls.}
\end{table}
\smallskip

\begin{table}[ht]
\begin{center}
\begin{tabular}{|l|c|c|c|}\hline
2017 elections & $\phi_1$ & $\theta_1$ & $JSD$ \\ \hline \hline
Conservative   & 0.8702   & 0.0202     & 0.0206 \\ \hline
Labour         & 0.8298   & 0.0317     & 0.0324 \\ \hline
\end{tabular}
\end{center}
\caption{\label{table:autocorr17-exp1} Parameters of the single exponential autocorrelation for the 2017 election polls.}
\end{table}
\begin{table}[ht]
\begin{center}
\begin{tabular}{|l|c|c|c|}\hline
2019 elections & $\phi_1$ & $\theta_1$ & $JSD$ \\ \hline \hline
Conservative   & 0.9775   & 0.0094     & 0.0043 \\ \hline
Labour         & 0.9592   & 0.0078     & 0.0056 \\ \hline
\end{tabular}
\end{center}
\caption{\label{table:autocorr19-exp1} Parameters of the single exponential autocorrelation for the 2019 election polls.}
\end{table}
\smallskip

We now turn our attention to the widely used {\em mean absolute error} (MAE) and {\em root mean square error} (RMSE) evaluation metrics \cite{CHAI14}, in order to directly estimate the prediction of the actual result using the EM method.
The MAE is given by
\begin{equation}\label{eq:mae}
MAE = \frac{\sum_{j=1}^{m} | p_j - f |}{m},
\end{equation}
where $p_j$ is the proportion favouring a political party in the $j$th poll,
$f$ is the corresponding proportion of votes in the actual election, and $m$ is the number of polls.
The RMSE is given by
\begin{equation}\label{eq:rmse}
RMSE = \sqrt{\frac{\sum_{j=1}^{m} \left(p_j - f\right)^2}{m}},
\end{equation}
noting that it is at least as large as the MAE.

\smallskip

We use the first third of the polls for computing the initial model parameter values, $\alpha$ and $\lambda$, of the gamma distribution, and also the $\phi_i$ and rate parameters $\theta_i$ in (\ref{eq:autocorr-sum}), with $n=2$. For each of the remaining two thirds of the polls, we adjust the parameters and use the EM method to predict the next step in the time series.  We repeat this twenty times and take the average of the twenty predictions at each time step to get the average prediction, and also compute the prediction when we set ${\mathrm d} W_t^{(i)}$ to zero in (\ref{eq:cir_sumi}), which is what we would expect the average to converge to
when increasing the number of EM computations, effectively eliminating the random component of the SDE represented by the diffusion function. We then compare the average prediction to the actual result of the election.
We evaluate the accuracy of the predictions over the complete range using the MAE and RMSE.
For comparison purposes, we also computed the MAE and RMSE using the current poll (the most recent poll inspected by the prediction algorithm) as the predictor of the actual result; these are shown in Tables \ref{table:mae-rmse-20} and \ref{table:mae-rmse-wt} in the columns labelled MAE-polls and RMSE-polls. The columns labelled MAE-EM and RMSE-EM show the error values of the predictions made using the EM method. It can be seen from these that for the two parties in both 2017 and 2019 the EM method was a better predictor than the polls themselves, and that the results in both tables are comparable; the margin of improvement is greatest for the Conservatives in 2019.

\begin{table}[ht]
\begin{center}
\begin{tabular}{|l|c|c|c|c|}\hline
Party-Year/Metric & MAE-polls & RMSE-polls & MAE-EM       & RMSE-EM     \\ \hline \hline
Con 2017          & 0.0278    & 0.0348     & {\em 0.0245} & {\em 0.0294} \\ \hline
Lab 2017          & 0.0988    & 0.1072     & {\em 0.0951} & {\em 0.0981} \\ \hline
Con 2019          & 0.0719    & 0.0949     & {\em 0.0644} & {\em 0.0853} \\ \hline
Lab 2019          & 0.0532    & 0.0618     & {\em 0.0491} & {\em 0.0568} \\ \hline
\end{tabular}
\end{center}
\caption{\label{table:mae-rmse-20} MAE and RMSE prediction errors for the 2017 and 2019 UK election results when averaging the predictions over twenty runs of the EM method. (The smaller error values are italicised.)}
\end{table}
\smallskip

\begin{table}[ht]
\begin{center}
\begin{tabular}{|l|c|c|c|c|}\hline
Party-Year/Metric & MAE-polls & RMSE-polls & MAE-EM       & RMSE-EM     \\ \hline \hline
Con 2017          & 0.0278    & 0.0348     & {\em 0.0250} & {\em 0.0296} \\ \hline
Lab 2017          & 0.0988    & 0.1072     & {\em 0.0951} & {\em 0.0977} \\ \hline
Con 2019          & 0.0719    & 0.0949     & {\em 0.0641} & {\em 0.0850} \\ \hline
Lab 2019          & 0.0532    & 0.0618     & {\em 0.0486} & {\em 0.0563} \\ \hline
\end{tabular}
\end{center}
\caption{\label{table:mae-rmse-wt} MAE and RMSE prediction errors for the 2017 and 2019 UK election results, when we set ${\mathrm d} W_t^{(i)}=0$. (The smaller error values are italicised.)}
\end{table}
\smallskip

We also counted the number of times the prediction using EM method was closer to the actual election result than was the prediction based on the current poll, and vice versa (cf. average ranks method \cite{BRAZ00});
the numbers are shown in Tables \ref{table:accuracy-20} and \ref{table:accuracy-wt} in the columns labelled Polls and EM.
The column labelled Total shows the total number of polls used, recalling that a third of the polls were used for computing the initial model parameters, while the column labelled Improvement shows the improvement percentage of the EM method prediction over using the polls themselves as predictors of the final result. These show a similar pattern to the prediction errors in Tables \ref{table:mae-rmse-20} and \ref{table:mae-rmse-wt}, i.e. in all cases the EM method is more accurate than using the polls themselves. The improvement is most notable for the Conservatives in 2019, and that for Labour in 2017 also stands out. We note that, apart from the improvement for the Conservatives in 2017, the results in Table~\ref{table:accuracy-20} are dominated by those in Table~\ref{table:accuracy-wt}.

\begin{table}[ht]
\begin{center}
\begin{tabular}{|l|c|c|c|c|}\hline
Party-Year & Polls & EM  & Total & Improvement \\ \hline \hline
Con 2017   & 79    & 90  & 169  & 6.51\%       \\ \hline
Lab 2017   & 72    & 97  & 169  & 14.79\%      \\ \hline
Con 2019   & 149   & 230 & 379  & 21.37\%      \\ \hline
Lab 2019   & 173   & 206 & 379  & 8.71\%       \\ \hline
\end{tabular}
\end{center}
\caption{\label{table:accuracy-20} Comparison of the number of times the closer prediction was based on either the current poll or using the EM method, when averaging the EM method predictions over twenty runs.}
\end{table}

\begin{table}[ht]
\begin{center}
\begin{tabular}{|l|c|c|c|c|}\hline
Party-Year & Polls & EM  & Total & Improvement \\ \hline \hline
Con 2017   & 80    & 89  & 169  & 5.33\%       \\ \hline
Lab 2017   & 70    & 99  & 169  & 17.16\%      \\ \hline
Con 2019   & 143   & 236 & 379  & 24.54\%      \\ \hline
Lab 2019   & 165   & 214 & 379  & 12.93\%       \\ \hline
\end{tabular}
\end{center}
\caption{\label{table:accuracy-wt} Comparison of the number of times the closer prediction was based on either the current poll or using the EM method, when we set ${\mathrm d} W_t^{(i)}=0$.}
\end{table}

\section{Concluding remarks}
\label{sec:conc}

We have proposed a generative SDE model to analyse the time series of opinion poll results leading up to an election. We have utilised a stochastic process that is the sum of CIR processes and has a stationary solution, where the marginal distribution of the time series is a gamma distribution. We provided empirical evidence that the model is a good fit to the polls leading up to the UK 2017 and 2019 elections. We also examined the predictive power of the model. We compared the errors in the predictions obtained using the EM method with those of the poll results themselves. We demonstrated that a model such as the one presented here may give better predictions of the actual election result than simply using the results of the polls.

\smallskip

One avenue for future work is to model the aggregation of distinct polls in terms of multiple time series models \cite{BRAN07}, and another is to generalise the model to deal multi-party systems may have more than two competing parties \cite{WALT15}.
It is also possible that the method we have presented using `mean-reverting' SDEs could augment an existing structural forecasting method,
or more generally be used as part of a toolkit for election prediction, resulting in  a synthetic \cite{LEWI16b} or hybrid \cite{PASE15} model that takes into account demographic data (cf. \cite{HANR18}).









\section*{Acknowledgements}

The authors would like to thank the reviewers for their constructive comments, which have helped us to improve the paper.

\end{document}